\documentclass{osa-article}

\journal{oe}

\articletype{Research Article}
\begin{document}

\title{Super-resolution single-photon imaging at 8.2 kilometers}

\author{Zheng-Ping Li,\authormark{1,2} Xin Huang,\authormark{1,2} Peng-Yu Jiang,\authormark{1,2} Yu Hong,\authormark{1,2} Chao Yu,\authormark{1,2} Yuan Cao,\authormark{1,2} Jun Zhang,\authormark{1,2} Feihu Xu,\authormark{1,2,*} and Jian-Wei Pan\authormark{1,2}}

% \cite{abdalati2010icesat}
\address{\authormark{1}National Laboratory for Physical Sciences at Microscale and Department of Modern Physics, University of Science and Technology of China, Hefei 230026, China\\
\authormark{2}CAS Center for Excellence in Quantum Information and Quantum Physics, University of Science and Technology of China, Hefei 230026, China}

\email{\authormark{*}feihuxu@ustc.edu.cn} %% email address is required

%%%%%%%%%%%%%%%%%%% abstract %%%%%%%%%%%%%%%%
\begin{abstract}
Single-photon light detection and ranging (LiDAR), offering single-photon sensitivity and picosecond time resolution, has been widely adopted for active imaging applications. Long-range active imaging is a great challenge, because the spatial resolution degrades significantly with the imaging range due to the diffraction limit of the optics, and only weak echo signal photons can return but mixed with a strong background noise. Here we propose and demonstrate a photon-efficient LiDAR approach that can achieve sub-Rayleigh resolution imaging over long ranges. This approach exploits fine sub-pixel scanning and a deconvolution algorithm tailored to this long-range application. Using this approach, we experimentally demonstrated active three-dimensional (3D) single-photon imaging by recognizing different postures of a mannequin model at a stand-off distance of 8.2 km in both daylight and night. The observed spatial (transversal) resolution is $\sim$5.5 cm at 8.2 km, which is about twice of the system's resolution. This also beats the optical system's Rayleigh criterion. The results are valuable for geosciences and target recognition over long ranges.
\end{abstract}

%%%%%%%%%%%%%%%%%%%%%%%%%%  body  %%%%%%%%%%%%%%%%%%%%%%%%%%
\section{Introduction}
Single-photon detection has become a well-established technique for the detection of weak optical signals. By exploiting this technique, single-photon light detection and ranging (LiDAR) offers excellent sensitivity, low noise and high time resolution~\cite{buller2007ranging}. Particularly, it can be used to remotely acquire three-dimensional (3D) shapes by making precise measurements of time-of-flight information, which has found applications in several scenarios, including geosciences, architecture, and defense.

Long-range and high-resolution active imaging is highly demanding for widespread applications, such as object detection, recognition and identification. When the imaging distance comes to tens of kilometers or more, very few photons can return to the detection system. Single-photon LiDAR, sensitive to the echo signal level as weak as a single photon, becomes an outstanding candidate. Tremendous efforts have been devoted to the developments of single-photon LiDAR for long-range active imaging \cite{mccarthy2009long,mccarthy2013kilometer,villa2014cmos,zhou2015few,li2017multi,pawlikowska2017single,ren2018high,chan2019long}. Single-photon 3D imaging at up to 45-km range has been reported lately \cite{li2019single}. Furthermore, computational imaging algorithms have seen remarkable progress to process the single-photon data efficiently~\cite{altmann2018quantum}. High-quality 3D structure has been demonstrated in the laboratory environment by an active imager detecting only one photon per pixel (PPP), based on the approaches of first-photon imaging~\cite{kirmani2013first}, pseudo-array \cite{shin2015photon,altmann2016lidar}, single-photon camera \cite{shin2016photon}, unmixing signal/noise \cite{rapp2017few} and machine learning \cite{lindell2018single}. These algorithms have the potential to improve the imaging range and quality significantly.

In long-range imaging, an important feature is the spatial (transversal) resolution. For a standard imaging system, the resolution is normally described as the angle determined by the diffraction limit, i.e., the aperture of the imaging system. This is also applied to single-photon LiDAR. An optimized scheme in single-photon LiDAR is to match the FoV of the detector and the point spread function (PSF) resulting from the transmission effect in the source-to-scene path (e.g., matching the diffraction limit and the aberration of transmission system)~\cite{mccarthy2013kilometer,pawlikowska2017single,li2019single}. Nonetheless, when the distance is in kilometers-long ranges, the spatial resolution decreases severely due to the divergence of the laser beam or the FoV of the receiver. Only matching the FoV and the PSF is not possible to increase the imaging resolution. To improve the resolution, an alternative approach is the fine sub-pixel scanning \cite{ur1992improved,park2003super,farsiu2004advances,sun2016improving}. The fine sub-pixel scanning precisely shifts the imager below the pixel scale to capture a series of low-resolution images and produces an image with higher resolution by combining multiple low-resolution images based on computational approaches. It can overcome the inherent spatial resolution limitation of the imaging system\cite{choi2004super,jiang2019scan}. To obtain different looks at the same scene, some relative scene motions must exist from frame sequences. In LiDAR system, setting the inter-pixel scanning space much smaller than the size of the receiver FoV can realize the fine sub-pixel scanning.

In this work, we demonstrate an effective super-resolution method to enhance the resolution of the long-range image captured by the single-photon LiDAR. Our method includes a sub-pixel scanning scheme in hardware and a matched 3D deconvolutional algorithm in software. The implementation of a high-efficiency, low-noise single-photon LiDAR system operating at the telecom wavelength of 1550 nm is presented. The performance of our super-resolution method is verified by both numerical simulations and outdoor experiments. In simulations, we show that our proposed method has the capability of achieving sub-Rayleigh resolution under light levels as low as $\sim$1 photon per pixel (PPP). In the outdoor experiments, we achieved super-resolution single-photon 3D imaging at long ranges up to 8.2 km in an urban environment. The experimental results demonstrate an adequate resolution of $\sim$5.5 cm to distinguish different postures of a mannequin model at a stand-off distance of 8.2 km, while conventional approaches fail to do so due to the limited resolution. The achieved resolution beats about twice of the system's resolution of $\sim$11.1 cm (or angle resolution of $\sim$13.5 $\mu rad$).

\section{Photon-efficient sub-pixel scanning approach}

The resolution of a standard imaging system is limited by its diffraction limit, which is mainly determined by the aperture of the system. In general, to balance the resolution and the detection efficiency, the field of view (FoV) of a single detection pixel is typically set to match this diffraction limit (Airy disk diameter). For single-photon LiDARs, an efficient way is to match the pixel size of the receiver with the point spread function (PSF) of the transmitter. In this case, the angle resolution of a LiDAR system $\Delta \theta$ can be described by,
\begin{equation} \label{E0}
\Delta \theta =\frac{2.44 \lambda}{D}.
\end{equation}
$\Delta \theta$ is determined by the operating wavelength $\lambda$, the diameter $D$ of the circular aperture for the aberration-free lens and the focal length $f$. Here 2.44 is the coefficient of the Airy pattern. Taking our single-photon LiDAR system as an example, the operating wavelength is $\lambda$ = 1550 nm and the diameter of the object lens is $D$ = 279 mm; the resolution of the system is $\Delta \theta$ = 13.5 $\mu rad$. At a distance of 8.2 km, this corresponds to the spatial resolution of $\sim$11.1 cm.

\begin{figure}[h!]
\centering\includegraphics[width=14cm]{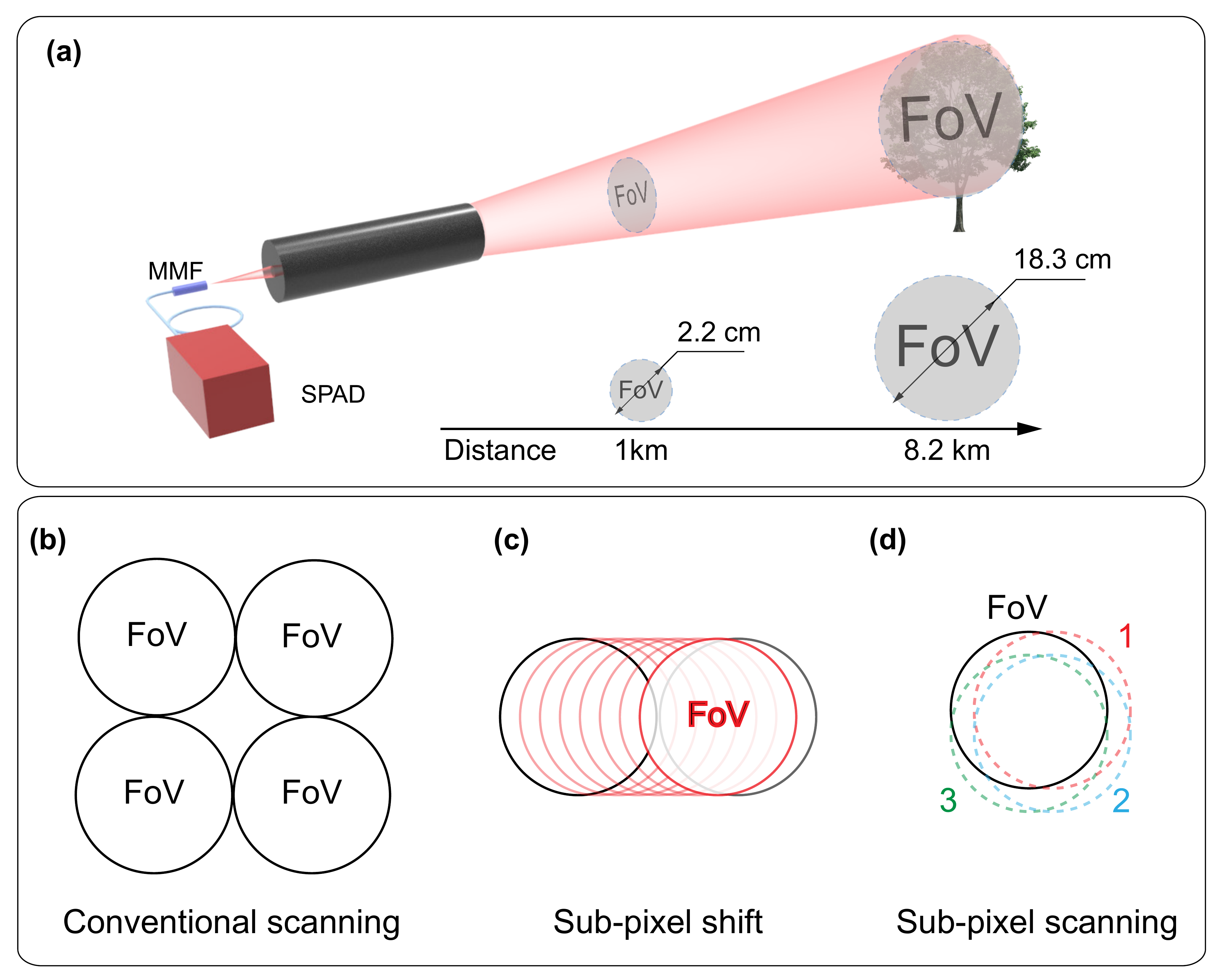}
\caption{\textbf{Schematic of the sub-pixel scanning method. (a).} Over long-range, even a very small FoV becomes a large patch projected on the target in far field. The spatial resolution deteriorates with the imaging distance. \textbf{(b).} Standard point-by-point scanning scheme. \textbf{(c).} The inter-pixel scanning space is set to be 1/8 FoV. \textbf{(d).} The sub-pixel scanning is performed in both $x$ and $y$ directions.}
\label{fig:sub}
\end{figure}

In long-range imaging, as shown in Fig.~\ref{fig:sub}(a), due to the divergence of the FoV,  even a small FoV that matches with the diffraction limit will turn into a large patch projected on the target, which deteriorates the resolution. This becomes notable for the imaging distance of kilometers long. To overcome this challenge, we propose to use the sub-pixel scanning method customized for single-photon LiDAR. Sub-pixel scanning scheme was initially proposed for ordinary digital cameras \cite{park2003super,farsiu2004advances}, where the imager is shifted at sub-pixel size scale to capture a series of low-resolution images. Then, an image with higher resolution can be computed from those low-resolution images. The sub-pixel displacements between these low-resolution images will offer frequency aliasing restraint, which is the essence of super-resolution reconstruction \cite{park2003super,farsiu2004advances}. The scheme has also been extended to single-pixel imaging~\cite{sun20133d} to improve the signal-to-background ratio (SBR) \cite{sun2016improving}.

In our coaxial single-photon LiDAR design, we set the FoV of the receiver slightly smaller than the divergence angle of the transmitting laser. This setting balances the need for high SBR and high collection efficiency, which has been widely adopted in long-range single-photon LiDAR systems \cite{mccarthy2013kilometer,pawlikowska2017single,li2019single}. To realize sub-pixel scanning, unlike the standard point-by-point scanning scheme [see Fig.~\ref{fig:sub}(b)], we set the inter-pixel scanning space smaller than the size of the receiver FoV. As an example, Fig.~\ref{fig:sub}(c) illustrates the inter-pixel spacing of 1/8 FoV. The inter-pixel shift is performed in both $x$ and $y$ directions [see Fig.~\ref{fig:sub}(d)]. After all pixels are scanned, a high-resolution image data will be computed from the combination of (8$\times $8) frames of low-resolution images. Note that the sub-Rayleigh resolution information offered by the sub-pixel scanning can be retrieved by a specific computational algorithm.

Another challenge in long range condition is the exceedingly weak echo signal. Recently, to deal with such extremely low-light levels, some photon-efficient algorithms were proposed~\cite{kirmani2013first,shin2015photon,altmann2016lidar,shin2016photon,rapp2017few}. However, most of them were only tested in the laboratory at a short distance. When it comes to the long-range situation, the divergence of the laser beam will lead to the problem of \emph{multiple returns} for each pixel \cite{shin2016computational,tachella20193d,li2019single,tachella2019bayesian,tachella2019real}. Previous algorithms can not retrieve the super-resolution information from the sub-pixel scanning data. Here we propose the 3D deconvolutional algorithm to solve the problem of multiple returns and to compute the high-resolution image from the sub-pixel scanning data with low signal levels at $\sim$1 PPP (see below).

\section{3D deconvolution algorithm}
To cooperate with the sub-pixel scanning approach, we develop an algorithm to perform the super-resolution photon-efficient imaging over long ranges. The advantages of our algorithm can be summarized in two aspects.

\begin{enumerate}
  \item We adopt a convolution forward model to take the fine scanning process into account and consider the issue of ``multiple returns per pixel" into consideration.
  \item We develop a \emph{3D deconvolution} method to retrieve the sub-pixel resolution information acquired from the fine scanning.
\end{enumerate}

\subsection{Forward model}
In our single-photon LiDAR system, we use a periodically pulsed laser to illuminate the target scene in a raster-scanned manner. The waveform of a single pulse is denoted by $w(t)$ with the full width at half maximum (FWHM) of $T_p$, and the repetition period is $T_r$. Now, we suppose the laser beam and the receiver are aimed at a scanning angle ($\theta_x$,$\theta_y$). Considering the divergence of the laser beam and the aperture of the receiver, a spatial kernel $g_{xy}$ is adopted to describe the spatial intensity distribution of the laser, and the FoV of the receiver projected on the scene. Based on the theory of the linear light transmission, the photon flux rate function $F(t;\theta_x,\theta_y)$ for the scanning angle ($\theta_x$,$\theta_y$) can be given as follows,
\begin{equation}
\label{E1}
\resizebox{.9\hsize}{!}{$F \left( t ; \theta _ { x } , \theta _ { y } \right) = \int _ { \theta _ { x } ^ { \prime } , \theta _ { y } ^ { \prime } \in FoV } g _ { x y } \left( \theta _ { x } - \theta _ { x } ^ { \prime } \right) r \left( \theta _ { x } ^ { \prime } , \theta _ { y } ^ { \prime } \right)\\ g _ { t } \left( t - 2 d \left( \theta _ { x } ^ { \prime } , \theta _ { y } ^ { \prime } \right) / c \right) d \theta _ { x } ^ { \prime } d \theta _ { y } ^ { \prime } + b$},
\end{equation}
for $t\in[0,T_r)$, where ($r(\theta_{x}^{ \prime }$,$\theta_y^{ \prime }$), $d(\theta_x^{ \prime }$,$\theta_y^{ \prime }$)) is the (reflectivity, depth) pair for the patch ($\theta_x^{ \prime }$,$\theta_y^{ \prime }$) on the scene; $c$ is the speed of light; $b$ denotes the background noise, and a temporal kernel $g_t$ describes the distribution of the system jitter. Here, we take the whole system jitter $g_t$ into account rather than the waveform of the laser pulse $w(t)$.

Specifically, the divergence of our transceiver system determines the FoV and the size of $g_{xy}$. In our experiment,  $g_{xy}$ is set to be a standard 2D Gaussian distribution, and its FWHM is set to be the size of the FoV (22.3 $\mu rad$); $g_t$ is set to be a standard 1D Gaussian distribution, and its FWHM is set to be 1 ns, which is equal to the system's timing jitter.

In practice, the continuous equation above has a discrete representation. We construct a 3D matrix $\mathbf{RD}$ whose $(i,j)th$ element is a vector with only one nonzero entry to describe the (reflectivity, depth) pair for the target scene. The value and the index of this entry represent the reflectivity and the depth of the scene respectively. Also, let a spatiotemporal kernel $g$ be the outer product of $h_{xy}$ and $h_t$, and B denotes the background noise matrix. Combining the inhomogeneous Poisson photon-detection processing \cite{snyder2012random}, the photon histogram matrix $\mathbf{Y}$ can be written as,
\begin{equation} \label{E2}
\mathbf{Y}\sim \text{Poisson}(\mathbf{g}\ast \mathbf{RD}+\mathbf{B}),
\end{equation}
where $\ast$ denotes the convolution operator.

\subsection{Reconstruction algorithm}

After taking the ``multiple returns per pixel" into consideration in the forward model, we design a 3D deconvolutional algorithm to compute the sub-Rayleigh resolution information. To get the fine estimation of $\mathbf{RD}$  from the raw data $\mathbf{Y}$ acquired from single-photon LiDAR, we treat this inverse problem as a single optimization problem and developed a deconvolutional convex optimization algorithm based on the forward model to solve it.

Let $\mathbf{L_{RD}}(\mathbf{RD};\mathbf{Y},\mathbf{g},\mathbf{B})$ be the negative log-likelihood function of the $\mathbf{RD}$ derived from Eq.~\eqref{E2}. The inverse regularized convex problem can be described as,
\begin{equation}
\label{E3}
\begin{array} { l l } { \underset { \mathbf{RD} } { \operatorname { minimize } } } & { \mathbf{L_{RD}} ( \mathbf{RD} ; \mathbf{Y}, \mathbf{g}, \mathbf{B} ) + \beta \cdot penalty ( \mathbf{RD} ) } \\ { \text {subject to } } & { \mathbf{R D _ { i , j , k} } \geq 0 , \forall i , j , k } \end{array}.
\end{equation}
Here, the constraint $RD_{i,j,k}\geq 0$ comes from the non-negativity of the reflectivity, and the smoothing term uses the total variation (TV) constrains. It is worth mentioning that our reconstruction framework is not restricted to a particular choice of the regularizer $\beta$.

Our 3D deconvolutional program, employing sequential quadratic approximations to the log-likelihood objective function $\mathbf{L_{RD}}$, is modified from the SPIRAL-TAP solver \cite{harmany2010spiral}. One main modification is that our program is performed under the 3D spatiotemporal domain to match the 3D convolutional operator $\mathbf{g}$, while the original SPIRAL-TAP solver was applied to 2D domain only. Here the core parameter in our 3D deconvolutional program is the spatiotemporal kernel $\mathbf{g}$. To cooperate with the sub-pixel scanning scheme, the spatial size of $\mathbf{g}$ needs to be adjusted according to the size of the inter-pixel spacing. For example, suppose we set the inter-pixel spacing to the size of 1/2 FoV, the spatial size of $\mathbf{g}$ is determined to be 3$\times$3. Generally, if we set the inter-pixel spacing to 1/(2n) FoV, the spatial size of $\mathbf{g}$ should be (2n+1)$\times$(2n+1) for n is a positive integer.

\section{Numerical simulations}
\begin{figure}[h!]
\centering\includegraphics[width=12cm]{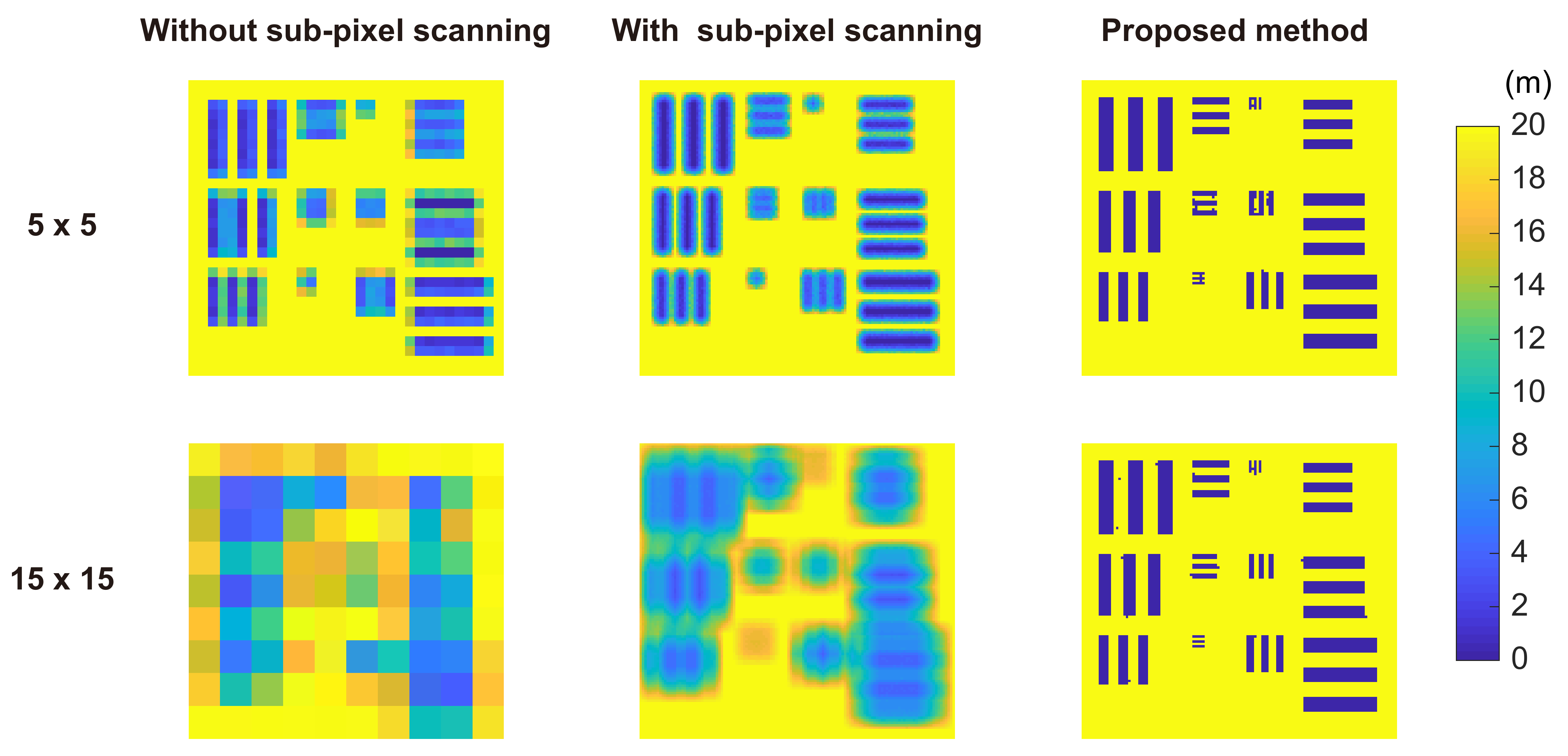}
\caption{\textbf{Ideal simulations on a resolution chart.} This simulation only takes the Poisson noise and system jitter into account. We simulated the FoV of two different sizes, 5$\times$5 and 15$\times$15, to illustrate the FoV for short and long distance. The first column shows the results without sub-pixel scanning. The second column shows the results with sub-pixel scanning and conventional pixelwise ML processing. The third column shows the results with sub-pixel scanning and our 3D deconvolutional processing. The step of the sub-pixel scanning is set to 1 pixel of the chart, i.e., the inter-pixel spacing is set to the size of the 1/4 FoV and 1/14 FoV respectively for the 5$\times$5 FoV and the 15$\times$15 FoV. From the results, we can get three inferences. First, when the imaging distance is becoming farther, the size of the FoV is becoming larger, and the resolution of the image becomes worse, as shown in the first column. Second, with the sub-pixel scanning scheme, the resolution becomes better, as shown in the second column. Third, our 3D deconvolutional algorithm substantially outperforms pixelwise ML.}
\label{fig:scanning}
\end{figure}

We provide numerical simulations to evaluate our sub-pixel scanning scheme. As shown in Fig.~\ref{fig:scanning}, we choose a resolution chart of size 120$\times$128 as our target scene. This chart consists of 6 squares of different sizes, each of which contains three bars and two spaces. In the largest square, both the bars and spaces are 6 pixels wide; in the smallest square, the features are 1 pixel wide. For simplicity, we set the background noise $\mathbf{B}$ to zero, i.e., only taking the Poisson noise and the system jitter into consideration. We simulated the FoV of two different sizes, 5$\times$5 and 15$\times$15, to illustrate the FoV for short and long distance. From Fig.~\ref{fig:scanning}, the results in the first column show that with the resolution of the image captured by the conventional FoV-by-FoV scanning method degrades with the expansion of the FoV. In the second column, we sub-pixel scan the scene with the smallest scanning step (shifting the FoV pixel by pixel on the chart), and compute the data with the conventional pixel-wise maximum likelihood (ML) method. In the last column, we compute the data with our 3D deconvolutional algorithm. Clearly, with our algorithm, the resolution of the reconstructed image has a substantial boost.

\begin{figure}[h!]
\centering\includegraphics[width=12cm]{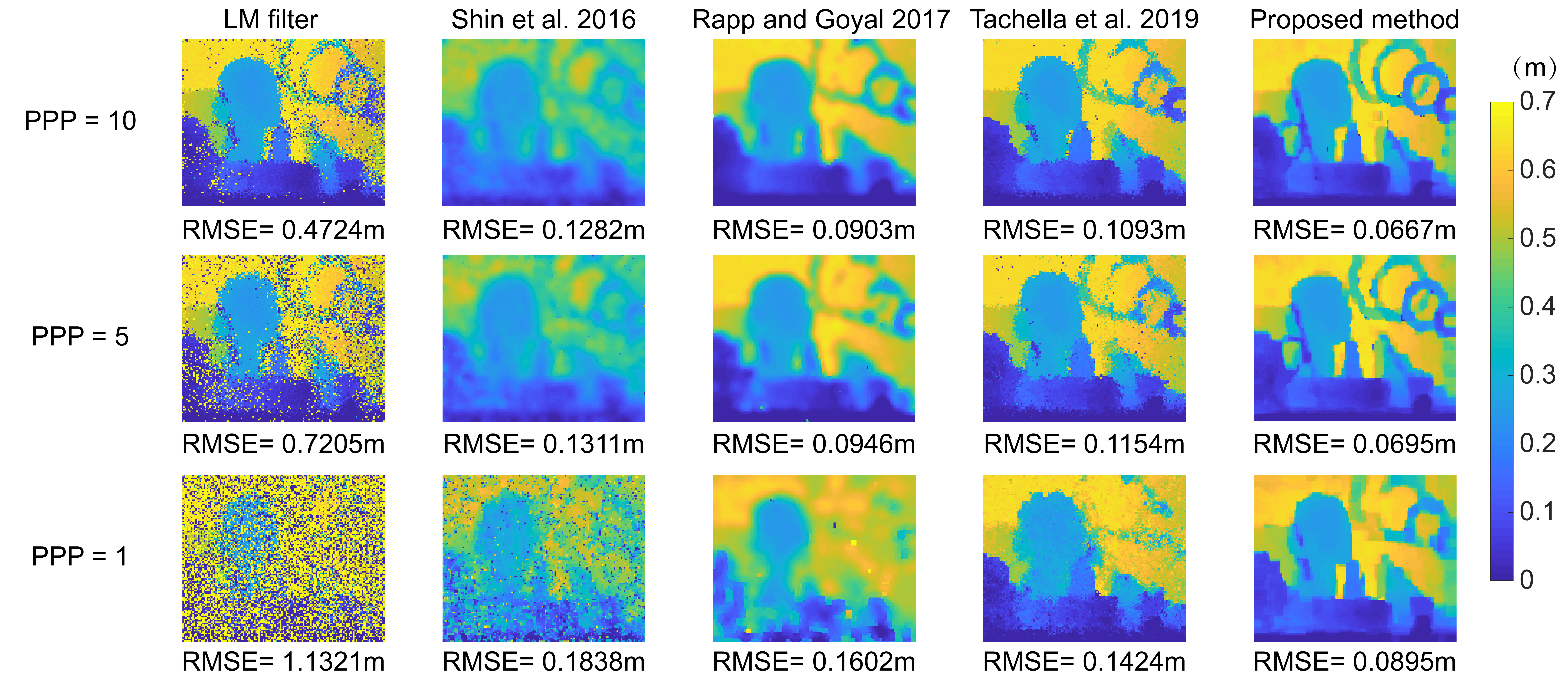}
\caption{\textbf{Low-light-level simulation.} The size of FoV is 9$\times$9, the SBR is 0.2, the average number of detected signal photons are 1, 5, 10 PPP. The sub-pixel scanning data is also processed by algorithms of log-match (LM) filter, Shin et al. 2016 \cite{shin2016photon}, Rapp and Goyal 2017 \cite{rapp2017few} and Tachella et al. 2019 \cite{tachella2019bayesian}. Quantitative results in terms of root mean square error (RMSE) are shown in the bottom of each figure. Clearly, our 3D deconvolutional algorithm has a smaller RMSE and superior performance to exhibit the details of the images.}
\label{fig:compare}
\end{figure}

To realize the high resolution in Fig.~\ref{fig:scanning}, a large number of signal photons are required. This is difficult to meet for practical long-range LiDAR system. In practice, only weak echo signal photons can return but mixed with a strong background noise. In Fig.~\ref{fig:compare}, we simulate the practical low-light conditions by setting the SBR ratio to 0.2 (within 100 ns time window similar to previous algorithms\cite{kirmani2013first,shin2016photon,rapp2017few}), and the inter-pixel spacing to 1/8 FoV. We choose a typical scene from the Middlebury dataset \cite{scharstein2007learning}. We simulated the results with the detected number of signal photons at 10, 5, 1 PPP, and compared the reconstruction results with state-of-the-art photon-efficient algorithms~\cite{shin2016photon,rapp2017few,tachella2019bayesian}. From the results shown in Fig.~\ref{fig:compare}, it is clear that our 3D deconvolutional method has a much better performance even under low-light levels and practical SBR conditions.

\section{Experimental setup}
\begin{figure}[h!]
\centering\includegraphics[width=12cm]{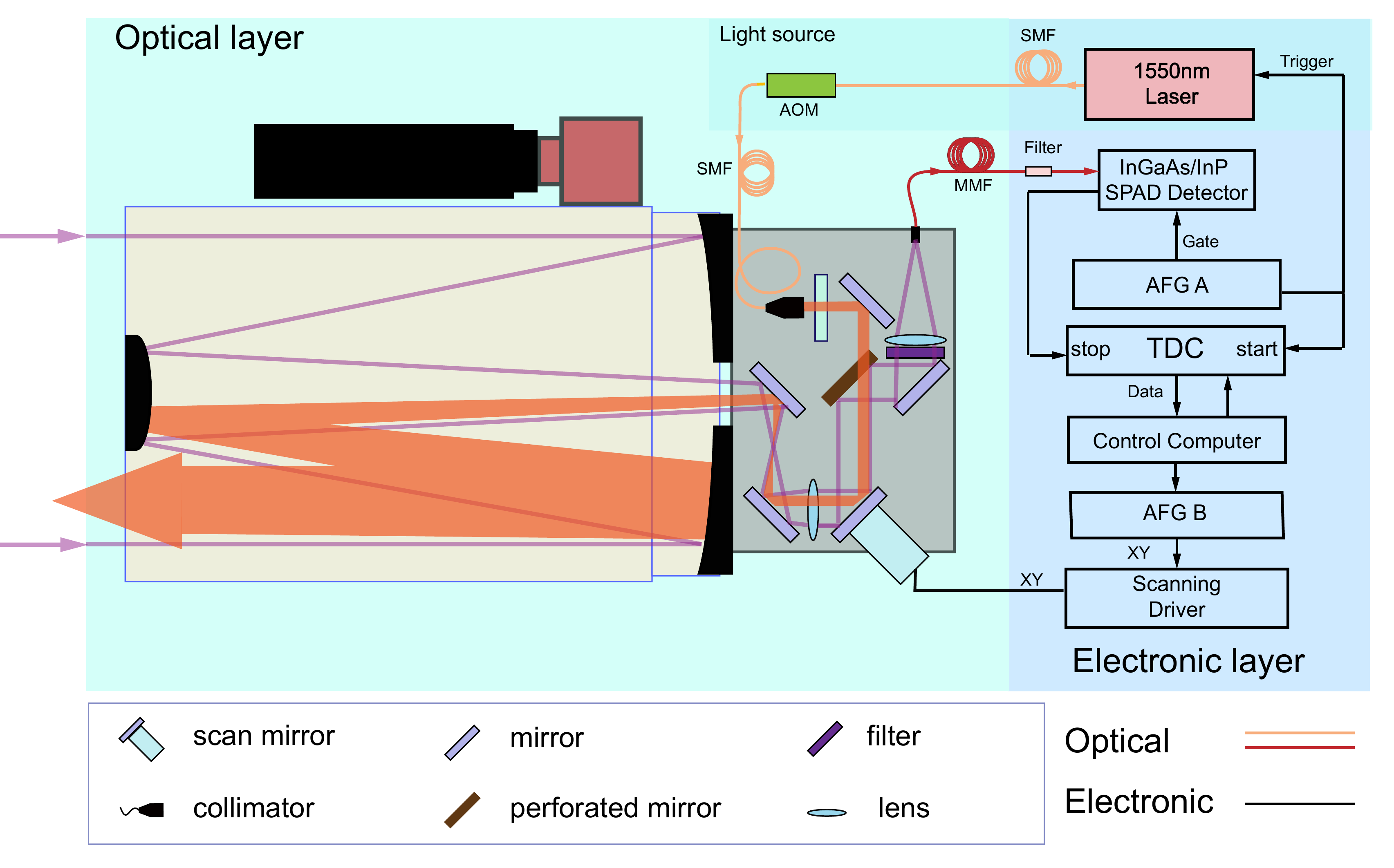}
\caption{\textbf{Schematic diagram of experimental setup.} AOM, acoustic-optical modulator; MMF, multimode fiber; PMF, polarization-maintaining fiber; AFG, Arbitrary Function Generator; TDC, time digital converter. A scanning mirror mounted on a piezo tip-tilt platform is used to steer the beam in both \emph{x} and \emph{y} axial directions. The optics is designed to near-diffraction limit. The transmitting divergence is about 35$\mu rad$. The receive fiber is a multi-mode fiber (MMF) with a core diameter of 62.5 $\mu m$ before the detector. It ensures a small FoV of 22.3 $\mu rad$. Both transmitter and receiver path pass through the same 28x expander coaxially which is the combination of the telescope (f = 2800 mm) and eyepiece (f = 100 mm). With this configuration, the receiver's FoV is slightly smaller than transmitter divergence. In addition, a standard camera (f=700 mm) was paraxially mounted on the telescope to provide a convenient direction and alignment aid for long distances.}
\label{fig:setup}
\end{figure}

\begin{table*}
 \centering
 \caption{\textbf{Summary of the system parameters}}
 \begin{tabular}{|c|c|}
  \hline\hline
  System Parameters &  Comment\\
  \hline\hline
  Wavelength & 1550 nm\\
  Distant & 8.2 km \\
  Scene & Mannequin model\\
  Size & 128 $\times$ 128 \\
  %System resolution & 13.5 $\mu rad$ ($\sim$ 11.1 cm at 8.2 km)\\
  FoV of the detector  & 22.3 $\mu rad$ ($\sim$ 18.3 cm at 8.2 km)\\
  Inter-pixel spacing & 1/8 FoV\\
  Spatial resolution & $\sim $ 5.5 cm at 8.2 km\\
  Detector &InGaAs/InP single photon detecter\\
  Detector Efficiency & 35\%\\
  Collection Fiber & 62.5 $\mu m$ MMF\\
  %Laser & --- & --- &\\
  Laser Repetition Rate & 100 kHz\\
  Laser Pulse Width & 0.5 ns\\
  Average Output Power & 120 mW\\
  Objective Lens & f=2800 mm, 279 mm diameter aperture\\
  Beam Scanning Mechanism &Coplanar dual-axis piezo tip/tilt platform \\
  Photon per pixel & 1-6\\
  SBR & 1/6-1/4 \\
  Algorithm & 3D deconvolutional \\
  \hline\hline
 \end{tabular} \label{table1}
 \end{table*}

Our experimental setup is shown in Fig.~\ref{fig:setup}. A summary of the system parameters is listed in Table~\ref{table1}. The optical transceiver system incorporates a commercial Cassegrain telescope with 279 mm aperture. We assembled optical components on a custom-built aluminum platform integrated with the telescope tube. The imaging system used a fiber laser operating at the wavelength of 1550 nm, which generates a 0.5 ns duration pulse at the repetition rate of 100 kHz. The 1550nm operating wavelength is used with the benefits of eye safety, reduction of the solar background and low atmospheric loss. The telecom brand components are readily available. The Commercial telescopes are typically coated with a 1550 nm coating to minimize the loss and back-reflections from the internal surfaces of the telescope. This is important in the case of the coaxial scanning LiDAR system with the amplified spontaneous emission (ASE) noise present. Furthermore, in particular, we have adopted an acousto-optical modulator (AOM) at the light source to filter the ASE component in time domain. The AOM acts as a high-speed switch. The emitted pulsed light passes through the AOM  with the transmissivity of about 60$\%$ , and the ASE light is isolated after the pulse in every repetition period. The ASE noise is almost eliminated, and after the moderation, the maximum transmitting power is about 120 mW.

The transceiver system was coaxial, allowing the area illuminated by the beam and the field of view (FoV) to remain matched while scanning. Precise sub-pixel scanning is implemented by a closed-loop piezo tip-tilt platform in both \emph{x} and \emph{y} axial directions. This coplanar dual-axis scanning scheme offers a capability of high-precision angle scanning. The arbitrary function generator (AFG) providing the scanning signal is set to precise voltage stepwise changes. A laser beam came out from a collimator and passed through a perforated mirror before expanded and transmitted from the telescope with a divergence angle of about 35 $\mu rad$. The returned photons reflected by the perforated mirror and passing through two wavelength filters (including a 1500-nm long-pass filter and a 9-nm bandpass filter), are collected by a focal lens into a filter based on multimode fiber (1.3 nm bandpass). Finally, the photons were detected by an InGaAs/InP single-photon avalanche diode detector (SPAD) \cite{Yu2017Fully}. The sensitive area of SPAD is a circle with a diameter of 25$\mu$m. When working in free-running mode, the dark count is $\sim$ 4.5 $counts/(s \cdot \mu m^2)$. In our experiment, the detector is enabled only 40\% of the time in each detection cycle, and the dark count is about 880 counts/s. In the transmitter, we employ a single mode fiber (SMF) and a fiber collimator with focal length f = 11 mm. Meanwhile, in the receiver path, we use a longer coupling lens with f = 100 mm. We employ a fiber filter (FF) based on multi-mode fiber (MMF) with a core diameter of 62.5 um before the detector. Both transmitter's and receiver's path coaxially pass through the same 28x expander which is the combination of the telescope (f = 2800 mm) and eyepiece (f = 100 mm). With this configuration, the receiver's FoV is slightly smaller than the divergence of the transmitter.

The FoV of our transceiver system is set near the diffraction limit, which is about 22.3 $\mu rad$. This is equivalent to a spatial resolution of $\sim$18 cm at a stand-off distance of 8.2 km. To beat the resolution limit, we perform an inter-pixel scanning. We set the inter-pixel scan spacing 1/8 receiver FoV (2.8 $\mu rad$). After scanning at 128$\times$128 points (pixels), a high-resolution image data containing (8 $\times$8) sub-scanning shifts is produced.

\section{Experiment results}
\begin{figure}[h!]
\centering\includegraphics[width=12cm]{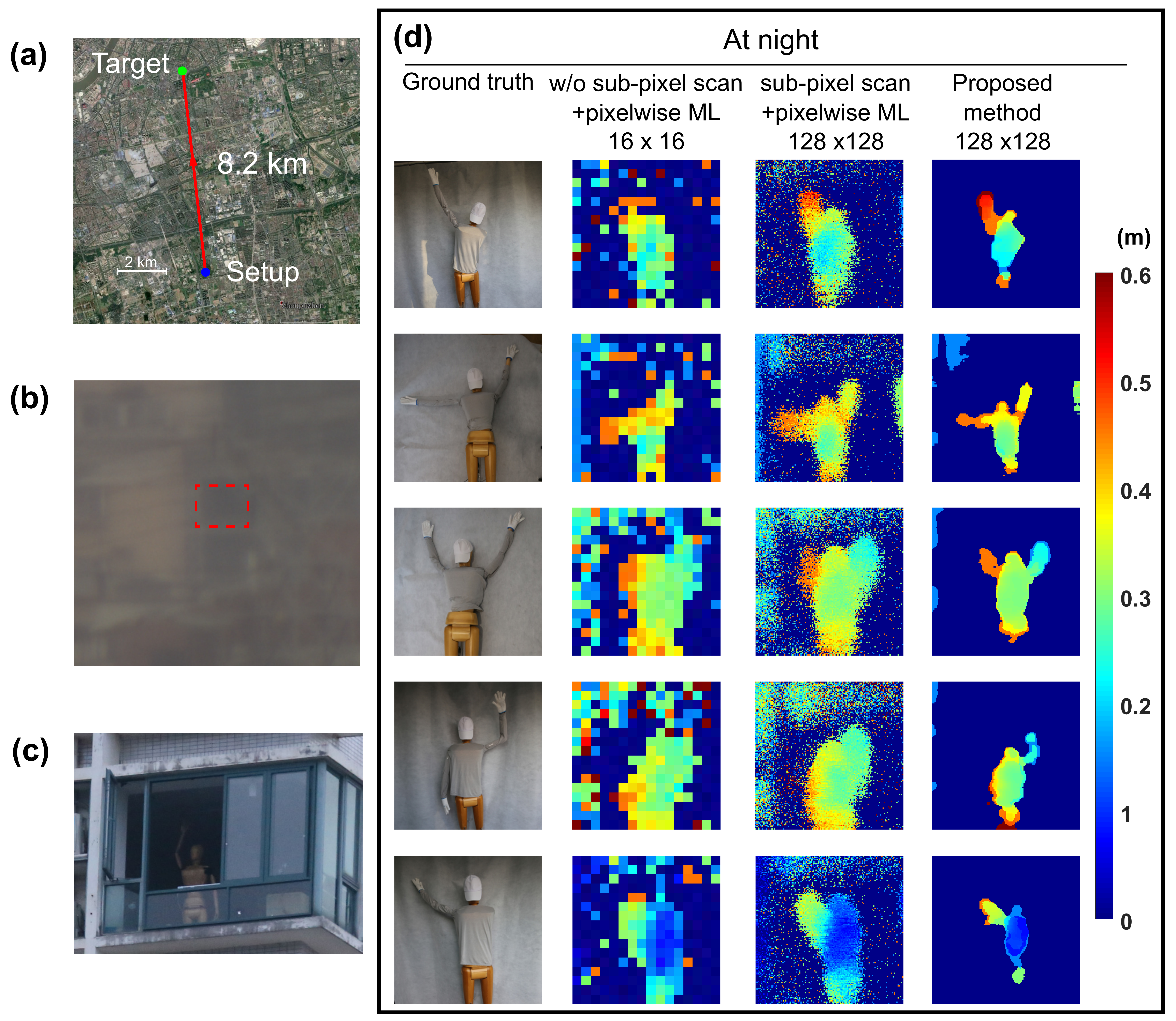}
\caption{\textbf{Experimental results.} \textbf{(a)}, Satellite image of the topology of the experiment in Shanghai city. \textbf{(b)}, Visible-band image of the target scene taken by a standard astronomical camera at a stand-off distance of 8.2 km. The red rectangle indicates the approximate LiDAR's FoV. \textbf{(c)}, Visible-band image of the target taken by a camera from a nearby building. A mannequin model with human size is placed in a room nearby the front window on 17-level floor of a tall building. \textbf{(d)} Imaging results of the different postures of the mannequin over 8.2 km range at night. The first column is the ground truth photon. The second column is the results without sub-pixel scanning. The third column is the reconstructed results with sub-pixel scanning and pixelwise ML. The forth column is the results with sub-pixel scanning and the 3D deconvolutional algorithm, where different postures of the human-size mannequin can be clearly recognized. The average signal photons of these results are $\sim$1 to 6 PPP and the SBR is $\sim$0.24.}
\label{fig:result}
\end{figure}

\begin{figure}[h!]
\centering\includegraphics[width=12cm]{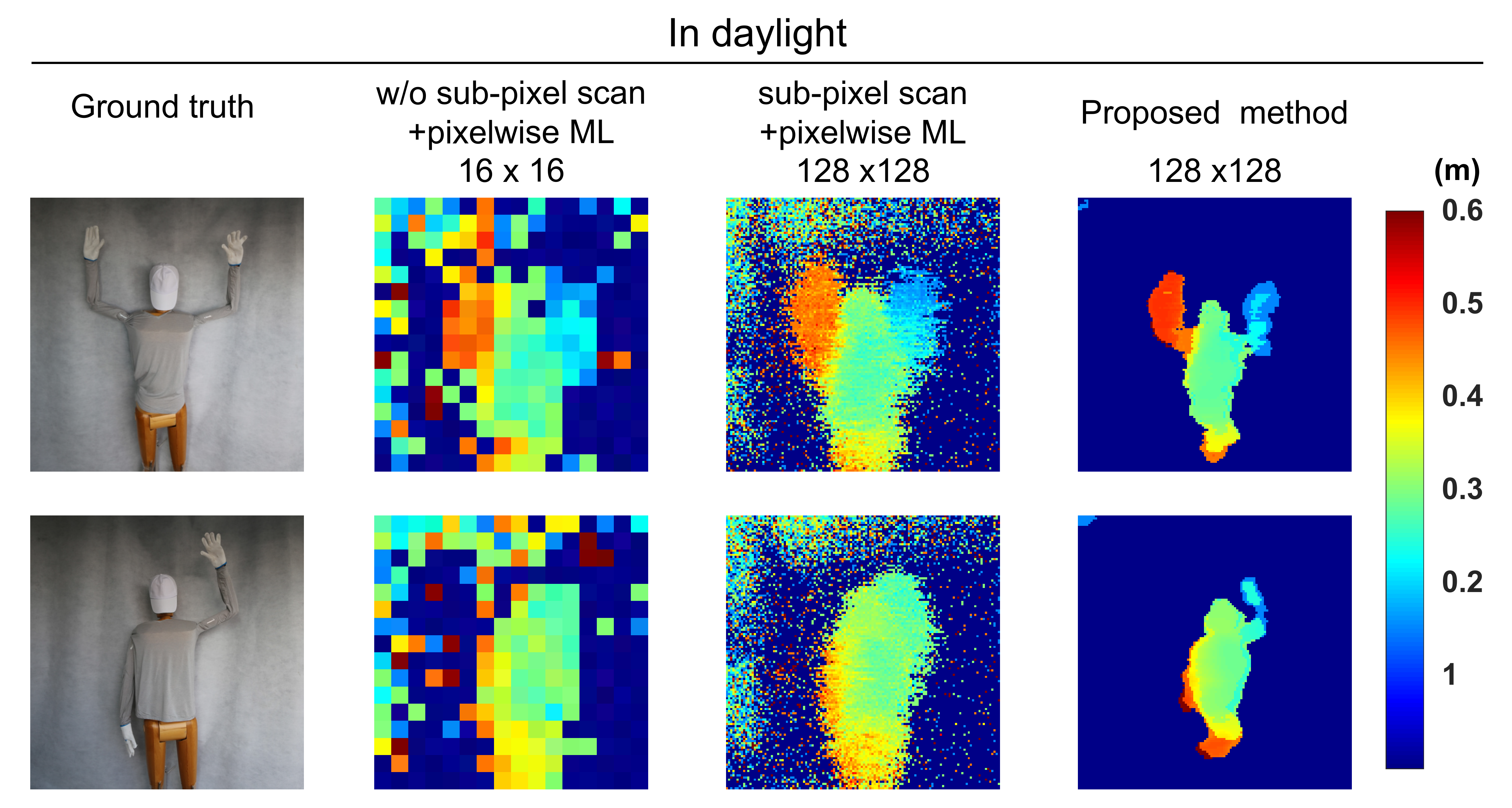}
\caption{\textbf{Experimental results in daylight.} Similar to Fig.~\ref{fig:result}, different postures of the mannequin at 8.2 km were taken by our single-photon LiDAR in daylight. The last column shows the results with sub-pixel scanning and the 3D deconvolutional algorithm. The average signal photons are $\sim$1 to 6 PPP and the SBR is $\sim$0.16.}
\label{fig:result2}
\end{figure}

As shown in Fig.~\ref{fig:result}(a), we tested and verified the capability of our super-resolution photon-efficient LiDAR system based on outdoor experiments in an urban environment of Shanghai city. Our aim is to recognize the postures of a mannequin model at a stand-off distance of 8.2 km in \emph{both} daylight and night. Before data acquisition, a photograph of the target was taken with a visible-band astronomical camera in daylight [see Fig.~\ref{fig:result}(b)]; the resulting visible-band image is substantially blurred because of the limited resolution of a standard camera and the urban's air turbulence in Shanghai. Fig.~\ref{fig:result}(c) shows a visible-band image taken by a commercial camera from a nearby building of a few meters.

In experiment, our single-photon LiDAR system operates with the maximum laser power of 120 mW and the acquisition time of 10 ms per pixel. According to the calibration of the signal and background noise in the acquired data, the number of signal photons is $\sim$1 to 6 PPP and the SBR is $\sim$0.16 to 0.24 (defined within 100-ns gate window similar to ref.~\cite{rapp2017few}). We successfully captured the fine depth maps of the mannequin postures in size of (128 $\times$ 128) pixels over 8.2 km in all time. The whole system jitter is measured to be 1 ns which is equivalent to 15 cm depth uncertainty.

The imaging results taken at night were shown in Fig.~\ref{fig:result}, including five different postures. The first column shows the ground-truth photos of the mannequin postures taken in the target room by a commercial camera. The second column shows the imaging results taken by the single-photon LiDAR without the sub-pixel scanning scheme. Basically, nothing can be outlined with such low resolutions. The third column shows the imaging results taken by the single-photon LiDAR with the sub-pixel scanning scheme and the pixelwise ML processing. It is obvious that with the finer scanning and the increase of the number of pixels, the resolution of the image becomes better. But, most of the postures still can not be distinguished. The last column shows our computed results on the sub-pixel scanning data with the 3D deconvolutional algorithm. This confirmed that the sub-pixel scanning scheme, together with the proposed reconstruction framework can substantially improve the imaging resolutions even with low signal photons on the level of $\sim$1 PPP. Different postures of the human-size mannequin can be clearly recognized and the head and arms can be seen from the depth imaging map. More importantly, the size of the arm of the mannequin model is $\sim$5.5 cm. This resolution beats about twice of the system's resolution of $\sim$11.1 cm (or angle resolution of 13.5 $\mu rad$).

We also performed the experiment in daylight and captured some other postures of the mannequin model. These results are exhibited in Fig.~\ref{fig:result2}. In contrast to the visible-band image in Fig.~\ref{fig:result}(b), our approach can successfully recognize different postures of the mannequin model over 8.2-km urban environment in daylight. These results demonstrate our system's all-time working ability. Certainly, due to the atmospheric influence such as turbulence in daylight, it is difficult for our LiDAR system to reach the highest resolution determined by the scanning spacing. Nevertheless, our sub-pixel scanning scheme greatly enhances the spatial resolution of the image over long ranges.

\section{Discussion}
We have proposed a super-resolution method for long-range single-photon LiDAR, including a sub-pixel scanning approach and a 3D deconvolutional algorithm. The superior performance of our method has been numerically and experimentally demonstrated. In experiment, depth profiles of the postures of human-size mannequin were clearly obtained at a stand-off distance of 8.2 km. We beat the diffraction limit of our LiDAR system, and achieved sub-Rayleigh resolution over kilometers range. The high-resolution results of different mannequin postures prove the effectiveness and practicability of our method. Additionally, the results captured in daylight and night show our system's adaptability for all-time applications. Overall, the high-resolution imaging under low light levels show the potential for target recognition and identification over long ranges.

\section*{Funding}
National Key Research and Development (R\&D) Plan of China (2018YFB0504300); National Natural Science Foundation of China (61771443); Anhui Initiative in Quantum Information Technologies (AHY140000); Chinese Academy of Sciences; Thousand Young Talent Program; Shanghai Science and Technology Development Funds (18JC1414700); Fundamental Research Funds for the Central Universities (WK2340000083).

\section*{Disclosures}
The authors declare no conflicts of interest.


\begin{thebibliography}{10}
\newcommand{\enquote}[1]{``#1''}

\bibitem{buller2007ranging}
G.~Buller and A.~Wallace, \enquote{Ranging and three-dimensional imaging using
  time-correlated single-photon counting and point-by-point acquisition,}
  {\protect\JournalTitle{IEEE Journal of Selected Topics in Quantum
  Electronics}} \textbf{13}, 1006--1015 (2007).

\bibitem{mccarthy2009long}
A.~McCarthy, R.~J. Collins, N.~J. Krichel, V.~Fern{\'a}ndez, A.~M. Wallace, and
  G.~S. Buller, \enquote{Long-range time-of-flight scanning sensor based on
  high-speed time-correlated single-photon counting,}
  {\protect\JournalTitle{Applied Optics}} \textbf{48}, 6241--6251 (2009).

\bibitem{mccarthy2013kilometer}
A.~McCarthy, X.~Ren, A.~Della~Frera, N.~R. Gemmell, N.~J. Krichel,
  C.~Scarcella, A.~Ruggeri, A.~Tosi, and G.~S. Buller, \enquote{Kilometer-range
  depth imaging at 1550 nm wavelength using an {InGaAs/InP} single-photon
  avalanche diode detector,} {\protect\JournalTitle{Optics Express}}
  \textbf{21}, 22098--22113 (2013).

\bibitem{villa2014cmos}
F.~Villa, R.~Lussana, D.~Bronzi, S.~Tisa, A.~Tosi, F.~Zappa, A.~Dalla~Mora,
  D.~Contini, D.~Durini, S.~Weyers, W.~ Brockherde, \enquote{{CMOS} imager with
  1024 spads and {TDCs} for single-photon timing and {3-D} time-of-flight,}
  {\protect\JournalTitle{IEEE Journal of Selected Topics in Quantum
  Electronics}} \textbf{20}, 364--373 (2014).

\bibitem{zhou2015few}
H.~Zhou, Y.~He, L.~You, S.~Chen, W.~Zhang, J.~Wu, Z.~Wang, and X.~Xie,
  \enquote{Few-photon imaging at 1550 nm using a low-timing-jitter
  superconducting nanowire single-photon detector,}
  {\protect\JournalTitle{Optics Express}} \textbf{23}, 14603--14611 (2015).

\bibitem{li2017multi}
Z.~Li, E.~Wu, C.~Pang, B.~Du, Y.~Tao, H.~Peng, H.~Zeng, and G.~Wu,
  \enquote{Multi-beam single-photon-counting three-dimensional imaging lidar,}
  {\protect\JournalTitle{Optics Express}} \textbf{25}, 10189--10195 (2017).

\bibitem{pawlikowska2017single}
A.~M. Pawlikowska, A.~Halimi, R.~A. Lamb, and G.~S. Buller,
  \enquote{Single-photon three-dimensional imaging at up to 10 kilometers
  range,} {\protect\JournalTitle{Optics Express}} \textbf{25}, 11919--11931
  (2017).

\bibitem{ren2018high}
X.~Ren, P.~W. Connolly, A.~Halimi, Y.~Altmann, S.~McLaughlin, I.~Gyongy, R.~K.
  Henderson, and G.~S. Buller, \enquote{High-resolution depth profiling using a
  range-gated {CMOS} spad quanta image sensor,} {\protect\JournalTitle{Optics
  Express}} \textbf{26}, 5541--5557 (2018).

\bibitem{chan2019long}
S.~Chan, A.~Halimi, F.~Zhu, I.~Gyongy, R.~K. Henderson, R.~Bowman,
  S.~McLaughlin, G.~S. Buller, and J.~Leach, \enquote{Long-range depth imaging
  using a single-photon detector array and non-local data fusion,}
  {\protect\JournalTitle{Scientific Reports}} \textbf{9}, 8075 (2019).

\bibitem{li2019single}
Z.-P. Li, X.~Huang, Y.~Cao, B.~Wang, Y.-H. Li, W.~Jin, C.~Yu, J.~Zhang,
  Q.~Zhang, C.-Z. ~Peng, F. ~Xu, J.-W. ~Pan, \enquote{Single-photon computational {3D}
  imaging at 45 km,} {\protect\JournalTitle{arXiv preprint arXiv:1904.10341}}
  (2019).

\bibitem{altmann2018quantum}
Y.~Altmann, S.~McLaughlin, M.~J. Padgett, V.~K. Goyal, A.~O. Hero, and
  D.~Faccio, \enquote{Quantum-inspired computational imaging,}
  {\protect\JournalTitle{Science}} \textbf{361}, eaat2298 (2018).

\bibitem{kirmani2013first}
A.~Kirmani, D.~Venkatraman, D.~Shin, A.~Cola{\c{c}}o, F.~N. Wong, J.~H.
  Shapiro, and V.~K. Goyal, \enquote{First-photon imaging,}
  {\protect\JournalTitle{Science}} \textbf{343}, 58--61 (2014).

\bibitem{shin2015photon}
D.~Shin, A.~Kirmani, V.~K. Goyal, and J.~H. Shapiro, \enquote{Photon-efficient
  computational {3-D} and reflectivity imaging with single-photon detectors,}
  {\protect\JournalTitle{IEEE Transactions on Computational Imaging}} \textbf{1}, 112--125
  (2015).

\bibitem{altmann2016lidar}
Y.~Altmann, X.~Ren, A.~McCarthy, G.~S. Buller, and S.~McLaughlin,
  \enquote{Lidar waveform-based analysis of depth images constructed using
  sparse single-photon data,} {\protect\JournalTitle{IEEE Transactions on Computational Imaging}} \textbf{25}, 1935--1946 (2016).

\bibitem{shin2016photon}
D.~Shin, F.~Xu, D.~Venkatraman, R.~Lussana, F.~Villa, F.~Zappa, V.~K. Goyal,
  F.~N. Wong, and J.~H. Shapiro, \enquote{Photon-efficient imaging with a
  single-photon camera,} {\protect\JournalTitle{Nature Communications}}
  \textbf{7}, 12046 (2016).

\bibitem{rapp2017few}
J.~Rapp and V.~K. Goyal, \enquote{A few photons among many: Unmixing signal and
  noise for photon-efficient active imaging,} {\protect\JournalTitle{IEEE Transactions on Computational Imaging}} \textbf{3}, 445--459 (2017).

\bibitem{lindell2018single}
D.~B. Lindell, M.~O'Toole, and G.~Wetzstein, \enquote{Single-photon {3D}
  imaging with deep sensor fusion,} {\protect\JournalTitle{ACM Transactions on Graphics (TOG) (TOG)}} \textbf{37}, 113 (2018).

\bibitem{ur1992improved}
H.~Ur and D.~Gross, \enquote{Improved resolution from subpixel shifted pictures,} {\protect\JournalTitle{CVGIP: Graphical Models and Image Processing}} \textbf{54},
181--186 (1992).

\bibitem{park2003super}
S.~C. Park, M.~K. Park, and M.~G. Kang, \enquote{Super-resolution image
  reconstruction: a technical overview,} {\protect\JournalTitle{IEEE Signal
  Processing Magazine}} \textbf{20}, 21--36 (2003).

\bibitem{farsiu2004advances}
S.~Farsiu, D.~Robinson, M.~Elad, and P.~Milanfar, \enquote{Advances and
  challenges in super-resolution,} {\protect\JournalTitle{International Journal
  of Imaging Systems and Technology}} \textbf{14}, 47--57 (2004).


\bibitem{sun2016improving}
M.-J. Sun, M.~P. Edgar, D.~B. Phillips, G.~M. Gibson, and M.~J. Padgett,
  \enquote{Improving the signal-to-noise ratio of single-pixel imaging using
  digital microscanning,} {\protect\JournalTitle{Optics Express}} \textbf{24},
  10476--10485 (2016).

\bibitem{choi2004super}
E.~Choi, J.~Choi, and M. G.~Kang, \enquote{Super-resolution approach to overcome physical limitations of imaging sensors: An overview,} {\protect\JournalTitle{International Journal of Imaging Systems and Technology}} \textbf{14},
36--46 (2004).

\bibitem{jiang2019scan}
S.~Jiang, X.~Li, Z.~Zhang, W.~Jiang, Y.~Wang, G.~He, Y.~Wang, B.~Sun \enquote{Scan efficiency of structured illumination in iterative single pixel imaging,} {\protect\JournalTitle{Optics Express}} \textbf{27},
22499--22507 (2019).

\bibitem{sun20133d}
B.~Sun, M.~P. Edgar, R.~Bowman, L.~E. Vittert, S.~Welsh, A.~Bowman, and
  M.~Padgett, \enquote{{3D} computational imaging with single-pixel detectors,}
  {\protect\JournalTitle{Science}} \textbf{340}, 844--847 (2013).

\bibitem{shin2016computational}
D.~Shin, F.~Xu, F.~N. Wong, J.~H. Shapiro, and V.~K. Goyal,
  \enquote{Computational multi-depth single-photon imaging,}
  {\protect\JournalTitle{Optics Express}} \textbf{24}, 1873--1888 (2016).

\bibitem{tachella20193d}
J.~Tachella, Y.~Altmann, S.~McLaughlin, and J.-Y. Tourneret, \enquote{{3D}
  reconstruction using single-photon lidar data exploiting the widths of the
  returns,} in \emph{ICASSP 2019-2019 IEEE International Conference on
  Acoustics, Speech and Signal Processing (ICASSP),}  (IEEE, 2019), pp.
  7815--7819.

\bibitem{tachella2019bayesian}
J.~Tachella, Y.~Altmann, X.~Ren, A.~McCarthy, G.~S. Buller, S.~McLaughlin, and J.-Y. Tourneret, \enquote{Bayesian 3D reconstruction of complex scenes from single-photon lidar data,}{\protect\JournalTitle{SIAM Journal on Imaging Sciences}} \textbf{12}, 521--550 (2019).

\bibitem{tachella2019real}
J.~Tachella, Y.~Altmann, N.~Mellado, A.~McCarthy, R.~Tobin, G.S.~Buller, J.-Y. Tourneret, and S.~McLaughlin, \enquote{Real-time 3D reconstruction from single-photon lidar data using plug-and-play point cloud denoisers,}{\protect\JournalTitle{Nature Communications}} \textbf{10}, 1--6 (2019).


\bibitem{snyder2012random}
D.~L. Snyder and M.~I. Miller, \emph{Random point processes in time and space}
  (Springer Science \& Business Media, 2012).

\bibitem{harmany2010spiral}
Z.~T. Harmany, R.~F. Marcia, and R.~M. Willett, \enquote{This is {SPIRAL-TAP}:
  Sparse poisson intensity reconstruction algorithms--theory and practice,}
  {\protect\JournalTitle{IEEE Transactions on Image Processing}} \textbf{21},
  1084--1096 (2012).

\bibitem{scharstein2007learning}
D.~Scharstein and C.~Pal, \enquote{Learning conditional random fields for
  stereo,} in \emph{2007 IEEE Conference on Computer Vision and Pattern
  Recognition,}  (IEEE, 2007), pp. 1--8.

\bibitem{Yu2017Fully}
C.~Yu, M.~Shangguan, H.~Xia, J.~Zhang, X.~Dou, and J.~W. Pan, \enquote{Fully
  integrated free-running {InGaAs/InP} single-photon detector for accurate
  lidar applications,} {\protect\JournalTitle{Optics Express}} \textbf{25},
  14611--14620 (2017).

\end{thebibliography}
\end{document}